\documentclass{iopart}
\usepackage{graphics}

\begin{document}
\jl{4}
\title[EFT of two- and many-body forces]{Towards a consistent approach to
nuclear structure: EFT of two- and many-body forces}
\author{R. Machleidt\dag\ and D. R. Entem\ddag}
\address{\dag\ Department of Physics, University of Idaho, Moscow,
Idaho 83844, U. S. A.}
\address{\ddag\ Nuclear Physics Group, University of Salamanca, E-37008 Salamanca, Spain}

\begin{abstract}
We review the nuclear forces currently in use, i.~e., the high-precision
NN potentials of the 1990's and the nuclear two- and many-body forces
based upon chiral effective field theory (EFT). We argue that the EFT approach
is superior to any of the older schemes. Since accurate chiral forces are
available now, the stage is set for microscopic nuclear struture to move
into a new and exciting era.
\end{abstract}

\section{Introduction}
This workshop deals essentially with microscopic nuclear structure
which has the goal
to derive the properties of atomic
nuclei from the `elementary' forces between nucleons.
Thus, the input for microscopic nuclear structure calculations 
are the `basic' nuclear forces:
two-nucleon forces (2NF), three-nucleon forces (3NF), \ldots .
It is the purpose of this contribution to review the nuclear
forces that are presently in use. We can distinguish between
two groups of current forces: the so-called high-precision nucleon-nucleon (NN)
potentials that were developed in the previous decade and
NN potentials (plus associated 3NFs) based upon chiral perturbation theory.
Presently, the majority of nuclear structure calculations are still conducted
with high-precision potentials, but the interest in the chiral forces
is increasing. The future belongs to the chiral approach for reasons
that we will explain in sect.~3.

Diversity has always been a characteristic feature of the
NN potential market. Therefore, it is not possible to squeeze all current 
potentials into the above two categories. Examples for
such exceptional cases are the Moscow potential~\cite{Kuk02},
the highly nonlocal potentials developed by Doleschall
and coworkers~\cite{Dol03}, and the NN potentials from inverse
scattering by Shirokov {\it et al.}~\cite{Shi04}. 
All these potentials have recently been applied in nuclear structure
yielding remarkable results. Unfortunately, because of lack of space, 
I cannot further elaborate on them.

\section{Historical perspective: The high-precision NN potentials}

In 1993, the Nijmegen group~\cite{Sto93} published a phase shift analysis
which described the NN data below 350 MeV laboratory energy
available at the time
with the `perfect' $\chi^2$/datum of 1.0.
This raised the expectation that also NN potentials should
reproduce the NN data with similar precision. 
Note that even the best NN models of the 1980's, like the
Paris~\cite{Lac80} and the Bonn~\cite{MHE87} potentials,
fit the NN data typically with a $\chi^2$/datum $\approx 2$ or more.
To put microscopic nuclear structure theory to a reliable test,
one needs a perfect NN potential such that discrepancies in the
predictions cannot be blamed on a bad fit of the NN data.

So, the Nijmegen analysis triggered a feverish activity
among groups traditionally involved in potential
construction. The output was a new family/generation of 
NN potentials which eventually were dubbed
the {\it high-precision potentials}.
The research groups involved and the names of their new creations are,
in chronological order:
\begin{itemize}
\item
Nijmegen group~\cite{Sto94}: Nijm-I, Nijm-II, and Reid93 potentials.
\item
Argonne group~\cite{WSS95}: $V_{18}$ potential.
\item
Bonn group~\cite{MSS96,Mac01}: CD-Bonn potential.
\end{itemize}
All these potentials have in common that they are charge-dependent
which is crucial to obtain a good fit of the $pp$ {\it and} $np$ data.
Moreover, they all need about 40-50 parameters to reproduce
the 1992 Nijmegen NN data
base with a $\chi^2$/datum $\approx 1$.
Note, however, that since 1993 the $pp$ database has substantially expanded
and for the current database the $\chi^2$/datum produced by some
of these potentials is not so perfect anymore
(cf.\ table~1).

\begin{table}[t]
\caption{$\chi^2$/datum for the reproduction of the 1992 and 1999
NN databases below 350 MeV by
the Nijmegen phase shift analysis~\protect\cite{Sto93} 
and two high-precision potentials:
the CD-Bonn potential~\protect\cite{Mac01} 
and the Argonne $V_{18}$ potential~\protect\cite{WSS95}.}
\footnotesize
\begin{indented}
\item[]\begin{tabular}{lccc}
\br
                   
 & CD-Bonn 
 & Nijmegen 
 & Argonne 
\\

 & potential
 & PSA
 & $V_{18}$ pot.\
\\
\br 
\multicolumn{4}{c}{\bf proton-proton data} \\
1992 $pp$ database (1787 data) & 1.00 & 1.00 & 1.10 \\
After-1992 $pp$ data (1145 data) & 1.03 & 1.24 & 1.74 \\
1999 $pp$ database (2932 data) & 1.01 & 1.09 & 1.35 \\
\hline 
\multicolumn{4}{c}{\bf neutron-proton data} \\
1992 $np$ database (2514 data) & 1.03 & 0.99 & 1.08 \\
After-1992 $np$ data (544 data) & 0.99 & 0.99 & 1.02 \\
1999 $np$ database (3058 data) & 1.02 & 0.99 & 1.07 \\
\hline 
\multicolumn{4}{c}{\bf {\boldmath $pp$} and {\boldmath $np$} data} \\
1992 $NN$ database (4301 data) & 1.02 & 0.99 & 1.09 \\
1999 $NN$ database (5990 data) & 1.02 & 1.04 & 1.21 \\
\br
\end{tabular}
\end{indented}
\end{table}

Concerning the theoretical basis of these potential, 
one could say that they are all---more or less---constructed
`in the spirit of meson theory' (e.g., all potentials include
the one-pion-exchange (OPE) contribution). However, there are
considerable differences in the details leading to considerable
off-shell differences among the potentials.

The CD-Bonn potential uses
the full, original, nonlocal Feynman amplitude for OPE, 
while all other potentials apply local approximations.
As a consequence, the CD-Bonn potential
has a weaker tensor force as compared to all other potentials.
This is reflected in the predicted D-state probabilities of
the deuteron, $P_D$, which is a measure of the strength of the nuclear tensor force.
While CD-Bonn predicts $P_D=4.85$\%, the other potentials
yield $P_D= 5.7(1)$\%.
These differences in the strength of the tensor force lead to
considerable differences in nuclear structure predictions.
The CD-Bonn potentials predicts 8.00 MeV for the triton binding
energy, while the local potentials predict only 7.62 MeV.
We note that the recent highly nonlocal Doleschall
potential~\cite{Dol03} predicts $P_D=3.6$\% 
and reproduces the entire triton binding of 8.48 MeV
from the 2NF alone.

The OPE contribution to the nuclear force essentially takes care of the
long-range interaction and the tensor force.
In addition to this, all models must describe the intermediate
and short range interaction, for which very different
approaches are taken.
The CD-Bonn includes (besides the pion)
the vector mesons $\rho (769)$ and $\omega (783)$, 
and two scalar-isoscalar bosons, $\sigma$, 
using the full, nonlocal Feynman amplitudes for their exchanges. 
Thus, all components of the CD-Bonn are nonlocal and the off-shell
behavior is the original one as determined from
relativistic field theory.

The models Nijm-I and Nijm-II are based upon the Nijmegen78
potential~\cite{NRS78}
which is constructed from approximate one-boson-exchange (OBE) amplitudes.
Whereas Nijm-II uses the local approximations
for all OBE contributions, Nijm-I keeps some nonlocal
terms in the central force component (but the Nijm-I
tensor force is local).
However,
nonlocalities in the central force have only a very
moderate impact on nuclear structure.
If one wants to retain nonlocality, 
it is more important to keep the tensor force nonlocalities.

The Reid93~\cite{Sto94} and Argonne $V_{18}$~\cite{WSS95} potentials do not 
use meson-exchange for 
intermediate and short range; instead, a phenomenological parametrization
is chosen.
The Argonne $V_{18}$ uses local functions
of Woods-Saxon type, 
while Reid93 applies local Yukawa functions of multiples
of the pion mass, similar to the original Reid potential
of 1968~\cite{Rei68}.
At very short distances, the potentials are regularized 
either by 
exponential ($V_{18}$, Nijm-I, Nijm-II) or by dipole (Reid93)
form factors, which are all local functions.

Over the past ten years, the family of high-precision
potentials has provided a useful service to the community.
Practitioners in the field of microscopic nuclear structure
and exact few-body calculations could finally be sure
that their predictions, good or bad, had nothing to do
with a bad fit of the NN data. Moreover, since all potentials
reproduce the NN data base with the same 
$\chi^2$/datum $\approx 1$ (`phase-equivalent' potentials), 
the impact of off-shell differences between
potentials on nuclear structure predictions 
could be studied systematically and 
reliably~\cite{Glo96,Eng97,Pol98,Cau02}.

However, in spite of these great practical achievements, the
high-precision potentials cannot be the end of the story.
From a fundamental point of view, these potentials
are not satisfactory at all. To achieve the acclaimed accuracy,
the potentials use about 45 parameters. The underlying
one-boson-exchange model has only about a dozen
parameters~\cite{Mac89}. Additional parameters are introduced,
e.g., by using partial wave dependent coupling constants,
which is hard to justify by the underlying model.
The motto is simply: 
\begin{quote}
\it
If you want more more accuracy,\\
you have to use less theory.
\end{quote}
This is a slap in the face of theoretical physics.
Moreover, fundamental questions are unanswered,
like: 
\begin{itemize}
\item
What is the connection to the fundamental theory
of strong interactions, QCD? 
\item
Why is the 2NF so much stronger
than the 3NF?
\end{itemize}
In the next section, we will show how to overcome these problems.

\section{The EFT approach to nuclear forces}

Quantum chromodynamics (QCD), the generally
accepted theory of strong interactions,
is non-perturbative in the low-energy
regime characteristic for nuclear physics.
For many years, this fact was perceived as the great obstacle for
a derivation of the nuclear force from QCD---impossible to overcome
except by lattice QCD.
The effective field theory (EFT) concept has shown the
way out of this dilemma.
Notice that the QCD Lagrangian for massless up and down quarks
is chirally symmetric, i.~e., it is invariant under
global flavor 
$SU(2)_L \times SU(2)_R$ equivalent to
$SU(2)_V \times SU(2)_A$ (vector and axial vector)
transformations. The axial symmetry is spontaneously broken
as evidenced in the absence of parity doublets in the
low-mass hadron spectrum. This implies the existence 
of three massless Goldstone bosons which are identified
with the three pions ($\pi^\pm, \pi^0$).
The non-zero, but small, pion mass is a consequence of
the fact that the up and down quark masses are not
exactly zero (some small, but explicit symmetry breaking).
Thus, we arrive at a low-energy scenario 
that consists 
of pions and nucleons interacting via a force
governed by spontaneously broken approximate chiral
symmetry. 

The effective Lagrangian describing this scenario
is given by an infinite series
of terms with increasing number of derivatives and/or nucleon
fields, with the dependence of each term on the pion field
prescribed by the rules of broken chiral symmetry.
Applying this Lagrangian to NN scattering generates an unlimited
number of Feynman diagrams.
However, Weinberg showed~\cite{Wei90} that a systematic expansion
exists in terms of $(Q/\Lambda_\chi)^\nu$,
where $Q$ denotes a momentum or pion mass, 
$\Lambda_\chi \approx 1$ GeV is the chiral symmetry breaking
scale, and $\nu \geq 0$ (cf.\ figure 1).
This has become known as chiral perturbation theory ($\chi$PT).
For a given order $\nu$, the number of contributing terms is
finite and calculable; these terms are uniquely defined and
the prediction at each order is model-independent.
By going to higher orders, the amplitude can be calculated
to any desired accuracy.
Thus, the motto is now: 
\begin{quote}
\it
If you want more more accuracy,\\
you have to use more theory (more orders).
\end{quote}
This sounds more like what we expect from theoretical physics.
Moreover, we have connected with QCD.

\begin{figure}
\hspace*{2cm}
\scalebox{0.5}{\includegraphics{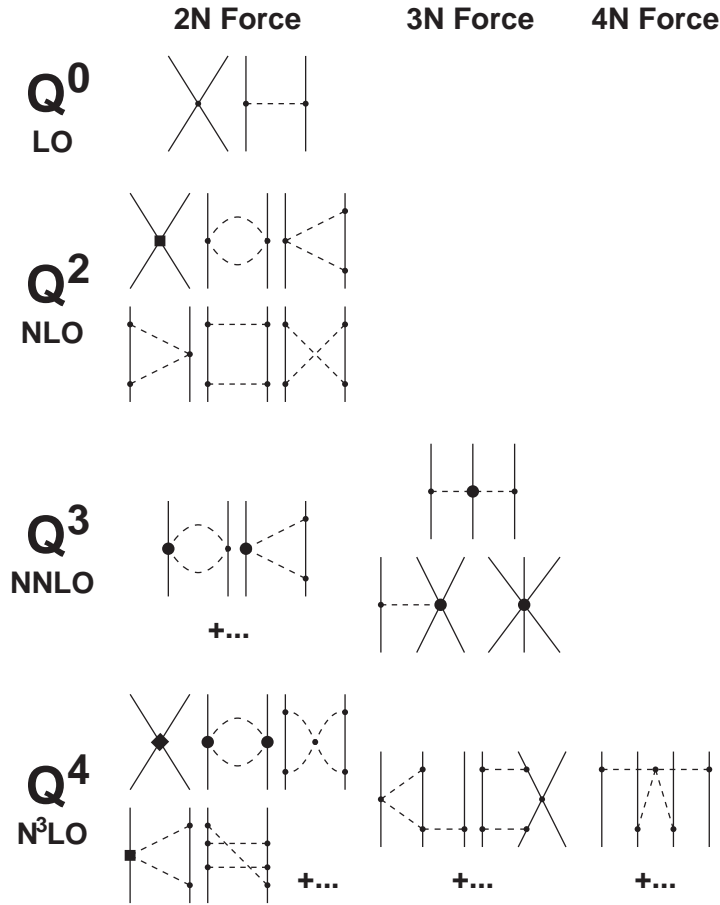}}
\vspace*{-0.5cm}
\caption{Hierarchy of nuclear forces in $\chi$PT. Solid lines
represent nucleons and dashed lines pions. Further explanations are
given in the text.}
\end{figure}

Following the first initiative by Weinberg \cite{Wei90}, pioneering
work was performed by Ord\'o\~nez, Ray, and
van Kolck \cite{ORK94,Kol99} who 
constructed a NN potential in coordinate space
based upon $\chi$PT at
next-to-next-to-leading order (NNLO; $\nu=3$).
The results were encouraging and
many researchers became attracted to the new field.
Kaiser, Brockmann, and Weise~\cite{KBW97} presented the first model-independent
prediction for the NN amplitudes of peripheral
partial waves at NNLO.
Epelbaum {\it et al.}~\cite{EGM98} developed the first momentum-space
NN potential at NNLO, and Entem and Machleidt~\cite{EM03} presented the first
potential at N$^3$LO.

In $\chi$PT, the NN amplitude is uniquely determined
by two classes of contributions: contact terms and pion-exchange
diagrams. There are two contacts of order $Q^0$ 
[${\cal O}(Q^0)$] represented by the four-nucleon graph
with a small-dot vertex shown in the first row of figure~1.
The corresponding graph in the second row, four nucleon legs
and a solid square, represent the 
seven contact terms of ${\cal O}(Q^2)$. 
Finally, at ${\cal O}(Q^4)$, we have 15 contact contributions
represented by a four-nucleon graph with a solid diamond.

Now, turning to the pion contributions:
At leading order [LO, ${\cal O}(Q^0)$, $\nu=0$], 
there is only the wellknown static one-pion exchange (OPE), second
diagram in the first row of figure~1.
Two-pion exchange (TPE) starts
at next-to-leading order (NLO, $\nu=2$) and all diagrams
of this leading-order two-pion exchange are shown.
Further TPE contributions occur in any higher order.
Of this sub-leading TPE, we show only
two representative diagrams at NNLO and three diagrams at N$^3$LO.
While TPE at NNLO was known for a while~\cite{ORK94,KBW97,EGM98},
TPE at N$^3$LO has been calculated only recently by
Kaiser~\cite{Kai01}. All $2\pi$ exchange diagrams/contributions up to
N$^3$LO are summarized in a pedagogical and systematic
fashion in Ref.~\cite{EM02} where 
the model-independent results for NN scattering in peripheral
partial waves are also shown. 

Finally, there is also three-pion exchange, which
shows up for the first time 
at N$^3$LO (two loops; one representative $3\pi$ diagram
is included in figure~1). 
In Ref.~\cite{Kai99},
it was demonstrated that the 3$\pi$ contribution at this order
is negligible. 

One important advantage of $\chi$PT is that it makes specific
predictions also for many-body forces. For a given order of $\chi$PT,
2NF, 3NF, \ldots are generated on the same footing (cf.\ figure~1). 
At LO, there are no 3NF, and
at next-to-leading order (NLO),
all 3NF terms cancel~\cite{Wei90,Kol94}. 
However, at NNLO and higher orders, well-defined, 
nonvanishing 3NF occur~\cite{Kol94,Epe02b}.
Since 3NF show up for the first time at NNLO, they are weak.
Four-nucleon forces (4NF) occur first at N$^3$LO and, therefore,
they are even weaker.

\begin{table}
\caption{Number of parameters used for the $np$ potential
in various approaches discussed in the text.}
\begin{indented}
\item[]\begin{tabular}{ccccc}
\br
                & Nijmegen & CD-Bonn & NLO        & N$^3$LO \\
                & PWA93    &``high$\;\;\;\;$& $Q^2$      & $Q^4$ \\
                & &$\;\;\;\;$precision''&$\;\;$(NNLO)$\;\;$&     \\
\br
$^1S_0$         & 3 & 4 & 2 & 4 \\
$^3S_1$         & 3 & 4 & 2 & 4 \\
\hline
$^3S_1$-$^3D_1$ & 2 & 2 & 1 & 3 \\
\hline
$^1P_1$         & 3 & 3 & 1 & 2 \\
$^3P_0$         & 3 & 2 & 1 & 2 \\
$^3P_1$         & 2 & 2 & 1 & 2 \\
$^3P_2$         & 3 & 3 & 1 & 2 \\
\hline
$^3P_2$-$^3F_2$ & 2 & 1 & 0 & 1 \\
\hline
$^1D_2$         & 2 & 3 & 0 & 1 \\
$^3D_1$         & 2 & 1 & 0 & 1 \\
$^3D_2$         & 2 & 2 & 0 & 1 \\
$^3D_3$         & 1 & 2 & 0 & 1 \\
\hline
$^3D_3$-$^3G_3$ & 1 & 0 & 0 & 0 \\
\hline
$^1F_3$         & 1 & 1 & 0 & 0 \\
$^3F_2$         & 1 & 2 & 0 & 0 \\
$^3F_3$         & 1 & 2 & 0 & 0 \\
$^3F_4$         & 2 & 1 & 0 & 0 \\
\hline
$^3F_4$-$^3H_4$ & 0 & 0 & 0 & 0 \\
\hline
$^1G_4$         & 1 & 0 & 0 & 0 \\
$^3G_3$         & 0 & 1 & 0 & 0 \\
$^3G_4$         & 0 & 1 & 0 & 0 \\
$^3G_5$         & 0 & 1 & 0 & 0 \\
\br
Total         & 35  & 38 & 9 & 24 \\
\br
\end{tabular}
\end{indented}
\end{table}

\section{Chiral NN potentials}

The two-nucleon system is non-perturbative as evidenced by the
presence of shallow bound states and large scattering lengths.
Weinberg~\cite{Wei90} showed that the strong enhancement of the
scattering amplitude arises from purely nucleonic intermediate
states. He therefore suggested to use perturbation theory to
calculate the NN potential and to apply this potential
in a scattering equation (Lippmann-Schwinger or Schr\"odinger 
equation) to obtain the NN amplitude. We follow 
this philosophy.

Chiral perturbation theory is a low-momentum expansion.
It is valid only for momenta $Q \ll \Lambda_\chi \approx 1$ GeV.
Therefore, when a potential is constructed, all expressions (contacts and
irreducible pion exchanges) are multiplied with a regulator function,
\begin{equation}
\exp\left[ 
-\left(\frac{p}{\Lambda}\right)^{2n}
-\left(\frac{p'}{\Lambda}\right)^{2n}
\right] \; ,
\end{equation}
where $p$ and $p'$ denote, respectively, the magnitudes
of the initial and final nucleon momenta in the center-of-mass
frame; and $\Lambda \ll \Lambda_\chi$. The exponent $2n$ is to be chosen
such that 
the regulator generates powers which are beyond
the order at which the calculation is conducted.

NN potentials based upon 
$\chi$PT at NNLO~\cite{EGM98,Epe02} are poor in quantitative terms; they
reproduce the NN data below 290 MeV
lab.\ energy with a $\chi^2$/datum of more than 20 (cf.\ tables 3 and 4, below).
As shown first by Entem and Machleidt~\cite{EM03},
one has to go to order N$^3$LO to obtain a NN potential
of acceptable accuracy.
Therefore, we will discuss now specifically the NN potential at N$^3$LO.

\begin{table}
\caption{
$\chi^2$/datum for the reproduction of the 1999 $np$ 
database below 290 MeV by various $np$ potentials.
($\Lambda=500$ MeV in all chiral potentials.)}
\begin{indented}
\item[]\begin{tabular}{cccccr}
\br
 Bin (MeV) 
 & \# of data 
 & N$^3$LO$^a$
 & NNLO$^b$
 & NLO$^b$ 
 & AV18$^c$
\\
\br 
0--100&1058&1.06&1.71&5.20&0.95\\
100--190&501&1.08&12.9&49.3&1.10\\
190--290&843&1.15&19.2&68.3&1.11\\
\hline
0--290&2402&1.10&10.1&36.2&1.04\\
\br
\end{tabular}
\\
\footnotesize
$^a$Reference~\cite{EM03}. \hspace{5mm}
$^b$Reference~\cite{Epe02}. \hspace{5mm}
$^c$Reference~\cite{WSS95}.
\end{indented}
\end{table}

\begin{table}[b]
\caption{
$\chi^2$/datum for the reproduction of the 1999 $pp$ 
database below 290 MeV by various $pp$ potentials.
($\Lambda=500$ MeV in all chiral potentials.)}
\begin{indented}
\item[]\begin{tabular}{cccccr}
\br
 Bin (MeV) 
 & \# of data
 & N$^3$LO$^a$
 & NNLO$^b$
 & NLO$^b$
 & AV18$^c$
\\
\br 
0--100&795&1.05&6.66&57.8&0.96\\
100--190&411&1.50&28.3&62.0&1.31\\
190--290&851&1.93&66.8&111.6&1.82\\
\hline
0--290&2057&1.50&35.4&80.1&1.38\\
\br
\end{tabular}
\\
\footnotesize
$^a$Reference~\cite{EM03}. \hspace{5mm}
$^b$See footnote~\cite{note3}. \hspace{5mm}
$^c$Reference~\cite{WSS95}.\\
\end{indented}
\end{table}

At N$^3$LO, there are 24 contact terms (24 parameters)
which contribute to the partial waves with $L\leq 2$.
In table~2, column `N$^3$LO/$Q^4$', we show how these
terms/parameters are distributed over the various partial waves.
For comparison, we also show the number of parameters
used in the Nijmegen partial wave analysis (PWA93)~\cite{Sto93}
and in the high-precision CD-Bonn potential~\cite{Mac01}.
The table reveals that, for $S$ and $P$ waves, the number of parameters
used in high-precision phenomenology and in EFT at N$^3$LO
are about the same.
Thus, the EFT approach provides retroactively a justification
for what the phenomenologist of the 1990's were doing.
At NLO and NNLO~\cite{foot}, the number of parameters is substantially
smaller than for PWA93 and CD-Bonn, which explains why
this order is insufficient for a quantitative potential.

For an accurate fit of the low-energy $pp$ and $np$ data, 
charge-dependence is important.
Charge-dependence up to next-to-leading order 
of the isospin-violation scheme 
(NL\O, in the notation of Ref.~\cite{WME01}) includes:
the pion mass difference in OPE and the Coulomb potential
in $pp$ scattering, which takes care of the L\O\/ contributions. 
At order NL\O\, we have pion mass difference in the NLO part of TPE,
$\pi\gamma$ exchange~\cite{Kol98}, and two charge-dependent
contact interactions of order $Q^0$ which make possible
an accurate fit of the three different $^1S_0$ scattering 
lengths, $a_{pp}$, $a_{nn}$, and $a_{np}$.

In the optimization procedure, we fit first phase shifts,
and then we refine the fit by minimizing the
$\chi^2$ obtained from a direct comparison with the data.
The $\chi^2/$datum for the fit of the $np$ data below
290 MeV is shown in table~3, and the corresponding one for $pp$
is given in table~4.
The $\chi^2$ tables show the quantitative improvement
of the NN interaction order by order in a dramatic way.
Even though there is considerable improvement when going from
NLO to NNLO, it is clearly seen that N$^3$LO is needed
to achieve an accuracy comparable
to  the phenomenological high-precision Argonne $V_{18}$
potential~\cite{WSS95}.
Note that proton-proton data
have, in general, smaller errors than $np$ data
which explains why the $pp$ $\chi^2$ are always larger.

The phase shifts for $np$ scattering below 300 MeV
lab.\ energy are displayed in figures 2.
What the $\chi^2$ tables revealed, can be seen graphically
in this figure. The $^2P_2$ phase shifts are a particularly
good example: NLO (dotted line) is clearly poor. NNLO
(dash-dotted line)
brings improvement and describes the data up to about
100 MeV. The difference between the NLO and NNLO
curves is representative for the uncertainty at NLO.
and, similarly, the difference between NNLO and N$^3$LO reflects
the uncertainty at NNLO.
Obviously, at N$^3$LO ($\Lambda=500$ MeV, thick solid line)
we have a good description up to 300 MeV. An idea
of the uncertainty at N$^3$LO can be obtained by varying
the cutoff parameter $\Lambda$. The thick dashed line
is N$^3$LO using $\Lambda=600$ MeV. 
In most cases, the latter two curves are not distinguishable
on the scale of the figures. Noticeable differences occur
only in $^1D_2$, $^3F_2$, and $\epsilon_2$ above 200 MeV.

\begin{figure}
\vspace*{-1.0cm}
\hspace*{-1.5cm}
\scalebox{0.4}{\includegraphics{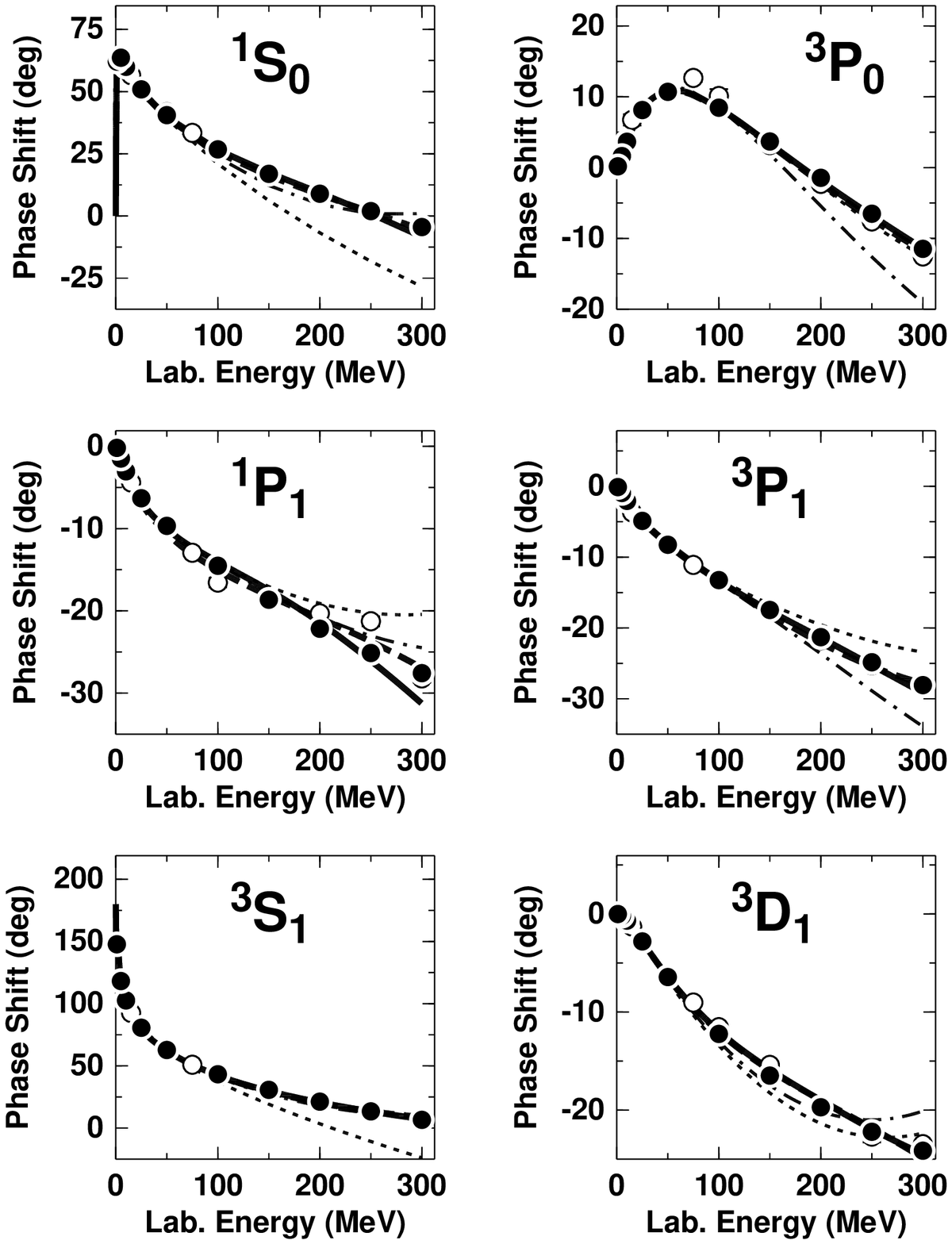}}
\hspace*{-2cm}
\scalebox{0.4}{\includegraphics{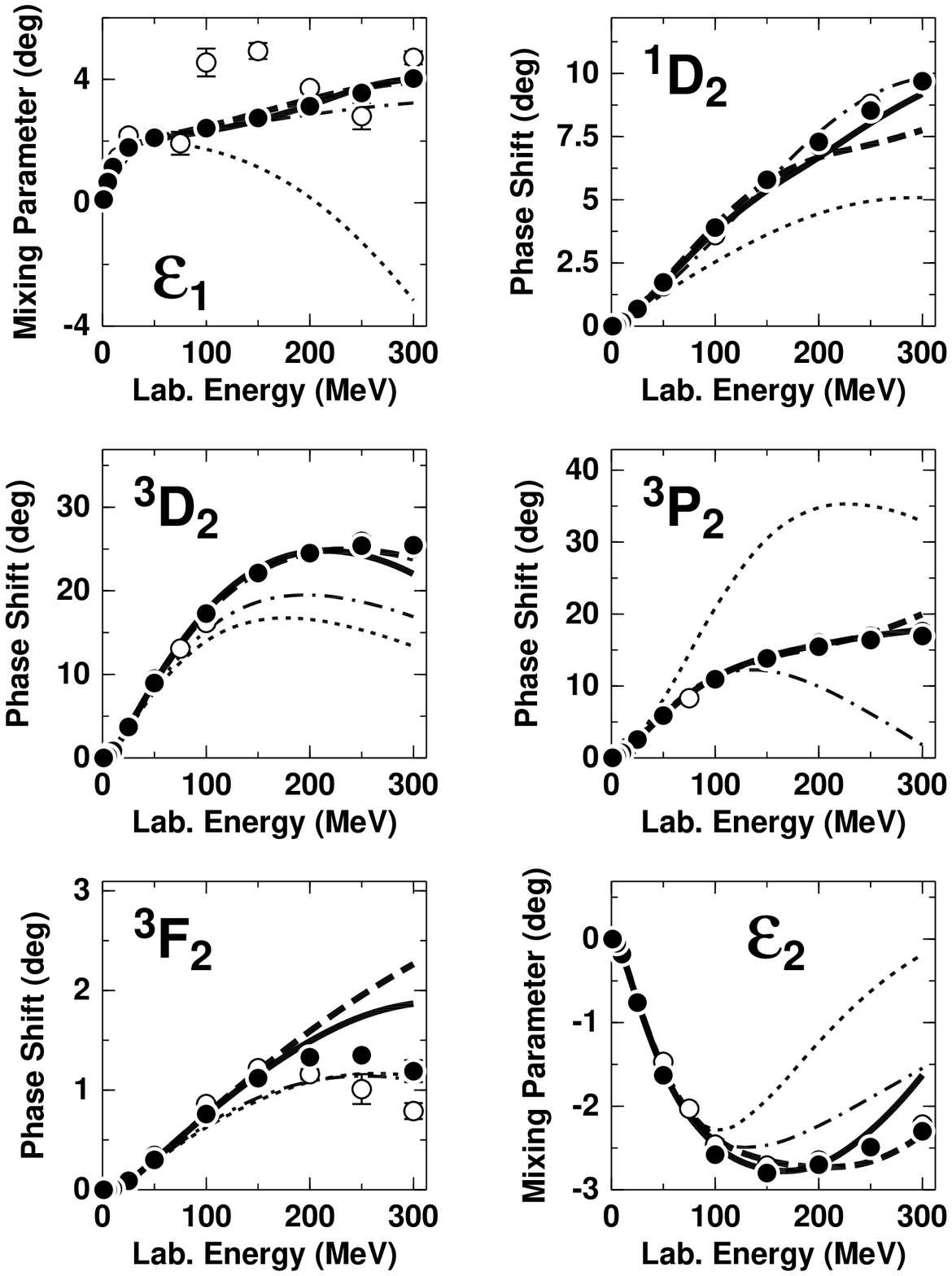}}
\vspace*{-2.0cm}
\caption{$np$ phase parameters below 300 MeV lab.\ energy for
partial waves with $J\leq 2$. The thick solid (dashed) line is the result
by Entem and Machleidt~\protect\cite{EM03}
at N$^3$LO using $\Lambda=500$ MeV ($\Lambda=600$ MeV).
The thin dotted and dash-dotted lines are the phase shifts at
NLO and NNLO, respectively, as obtained by 
Epelbaum {\it et al.}~\protect\cite{Epe02} using $\Lambda=500$ MeV.
The solid dots show the Nijmegen multienergy $np$ phase shift
analysis~\protect\cite{Sto93}, and the open circles are the GWU/VPI 
single-energy $np$ analysis SM99~\protect\cite{SM99}.}
\end{figure}

\section{Chiral three-nucleon forces}

As noted before, the first non-vanishing 3NF terms
occur at NNLO and are shown in figure~1
(row `$Q^3$/N$^2$LO', column `3N Force'). 
There are three diagrams: the TPE, OPE,
and 3N-contact interactions~\cite{Epe02b}.
The TPE 3N-potential is given by
\begin{equation}
V^{\rm 3NF}_{\rm TPE} = 
\left( \frac{g_A}{2f_\pi} \right)^2
\frac12 
\sum_{i \neq j \neq k}
\frac{
( \vec \sigma_i \cdot \vec q_i ) 
( \vec \sigma_j \cdot \vec q_j ) }{
( q^2_i + m^2_\pi )
( q^2_j + m^2_\pi ) } \;
F^{\alpha\beta}_{ijk} \;
\tau^\alpha_i \tau^\beta_j
\end{equation}
with $\vec q_i \equiv \vec{p_i}' - \vec p_i$, 
where 
$\vec p_i$ and $\vec{p_i}'$ are the initial
and final momenta of nucleon $i$, respectively, and
\begin{equation}
F^{\alpha\beta}_{ijk} = \delta^{\alpha\beta}
\left[ - \frac{4c_1 m^2_\pi}{f^2_\pi}
+ \frac{2c_3}{f^2_\pi} \; \vec q_i \cdot \vec q_j \right]
+ 
\frac{c_4}{f^2_\pi}  
\sum_{\gamma} 
\epsilon^{\alpha\beta\gamma} \;
\tau^\gamma_k \; \vec \sigma_k \cdot [ \vec q_i \times \vec q_j] \; .
\end{equation}  
The vertex involved in this 3NF term is the two-derivative
$\pi\pi NN$ vertex (large solid dot in figure~1) which we encountered
already  in the TPE contribution to the 2N potential at NNLO.
Thus, there are no new parameters and the contribution
is fixed by the LECs used in NN.
The OPE contribution is
\begin{equation}
V^{\rm 3NF}_{\rm OPE} = 
D \; \frac{g_A}{8f^2_\pi} 
\sum_{i \neq j \neq k}
\frac{\vec \sigma_j \cdot \vec q_j}{
 q^2_j + m^2_\pi }
( \mbox{\boldmath $\tau$}_i \cdot \mbox{\boldmath $\tau$}_j ) 
( \vec \sigma_i \cdot \vec q_j ) 
\end{equation}
and, finally, the 3N contact term reads
\begin{equation}
V^{\rm 3NF}_{\rm ct} = E \; \frac12
\sum_{j \neq k} 
 \mbox{\boldmath $\tau$}_j \cdot \mbox{\boldmath $\tau$}_k  \; .
\end{equation}
The last two 3NF terms involve two new vertices
(that do not occur in the 2N problem), namely,
the $\pi NNNN$ vertex with parameter $D$
and a $6N$ vertex with parameters $E$,
To pin them down, one needs two
3N observables. In reference~\cite{Epe02b},
the triton binding energy and the $nd$ doublet scattering
length $^2a_{nd}$ were used.
Once $D$ and $E$ are fixed, the results for other
3N, 4N, \ldots observables are predictions.
In reference~\cite{Nog04}, encouraging results were reported 
for $^6$Li. 
Concerning the famous `$A_y$ puzzle', the above 3NF terms
yield some improvement of the predcited $nd$ $A_y$, however,
the problem is not resolved.

One should note that there are additional 3NF terms at NNLO 
due to relativistic corrections ($1/M_N$ corrections) that
have not yet been included in any calculation.
However, there are all reasons to believe
that these contributions will be very small, probably
negligible.
It is more likely that the problem with the chiral 3NF
is analogous to the one with the chiral 2NF:
namely, NNLO is insufficient and for sufficient
accuracy one has to proceed to N$^3$LO.
Two 3NF terms at N$^3$LO are indicated in figure~1.
The N$^3$LO 3NF, which will probably not depend on
any new parameters, is presently under development.

\section{Conclusions}

The EFT approach to nuclear forces is superior to all earlier
nuclear force models. It represents a scheme that has an intimate
relationship with QCD and allows to calculate nuclear forces
to any desired accuracy. Moreover, nuclear two- and many-body
forces are generated on the same footing.

At N$^3$LO~\cite{EM03}, the accuracy is achieved that
is necessary and sufficient for reliable nuclear structure
calculations. First calculations applying the N$^3$LO
potential have produced promising 
results~\cite{Nog04,Kow04,DH04,Wlo05,Dea04,NC03,FNO04}.

The 3NF at NNLO is known~\cite{Epe02b} and has had
first successful applications~\cite{Nog04}. The
3NF and 4NF at N$^3$LO is presently under construction.

In summary, the stage is set for many years of exciting
nuclear structure research that is more consistent
than anything we had before.

\ack
This work was supported by the U.S. National Science
Foundation under Grant No.~PHY-0099444, by 
the Spanish  Ministerio de Ciencia y Tecnolog{\'\i}a 
under Contract No. BFM2001-3563, and the Junta de Castilla y Le\'on under 
Contract No. SA-104/04.


\begin{thebibliography}{99}
\bibitem{foot} Note that, due to parity, contact
terms appear only at even powers of $Q$ which is why
the contacts are the same for NLO and NNLO.
\end{thebibliography}
\end{document}